\documentstyle[epsfig]{mn2e}
\begin{document}

\title[Universal fitting formulae for baryon oscillation
  surveys]{Universal fitting formulae for baryon oscillation surveys}

\author[Chris Blake et al.]{\parbox[t]{\textwidth}{Chris
    Blake$^{1,}$\footnotemark, David Parkinson$^2$, Bruce
    Bassett$^{3,4}$, Karl Glazebrook$^5$, \\ Martin Kunz$^{2,6}$,
    Robert C.\ Nichol$^3$} \\ \\ $^1$ Department of Physics \&
  Astronomy, University of British Columbia, 6224 Agricultural Road,
  Vancouver, B.C., V6T 1Z1, Canada \\ $^2$ Astronomy Centre,
  University of Sussex, Brighton, BN1 9QH, U.K. \\ $^3$ Institute of
  Cosmology \& Gravitation, University of Portsmouth, Portsmouth, P01
  2EG, U.K. \\ $^4$ South African Astronomical Observatory, P.O.\ Box
  9, Observatory 7935, Cape Town, South Africa \\ $^5$ Department of
  Physics \& Astronomy, Johns Hopkins University, Baltimore, MD
  21218-2686, United States \\ $^6$ Department of Theoretical Physics,
  University of Geneva, 24 quai Ernest Ansermet, CH-1211 Geneva 4,
  Switzerland}

\maketitle

\begin{abstract}
The next generation of galaxy surveys will attempt to measure the
baryon oscillations in the clustering power spectrum with high
accuracy.  These oscillations encode a preferred scale which may be
used as a standard ruler to constrain cosmological parameters and dark
energy models.  In this paper we present simple analytical fitting
formulae for the accuracy with which the preferred scale may be
determined in the tangential and radial directions by future
spectroscopic and photometric galaxy redshift surveys.  We express
these accuracies as a function of survey parameters such as the
central redshift, volume, galaxy number density and (where applicable)
photometric redshift error.  These fitting formulae should greatly
increase the efficiency of optimizing future surveys, which requires
analysis of a potentially vast number of survey configurations and
cosmological models.  The formulae are calibrated using a grid of
Monte Carlo simulations, which are analyzed by dividing out the
overall shape of the power spectrum before fitting a simple decaying
sinusoid to the oscillations.  The fitting formulae reproduce the
simulation results with a fractional scatter of $7\%$ ($10\%$) in the
tangential (radial) directions over a wide range of input parameters.
We also indicate how sparse-sampling strategies may enhance the
effective survey area if the sampling scale is much smaller than the
projected baryon oscillation scale.
\end{abstract}
\begin{keywords}
large-scale structure of Universe -- cosmological parameters --
surveys
\end{keywords}

\section{Introduction}
\renewcommand{\thefootnote}{\fnsymbol{footnote}}
\setcounter{footnote}{1}
\footnotetext{E-mail: cab@astro.ubc.ca}

Baryon oscillations in the galaxy power spectrum have recently emerged
as a promising standard ruler for cosmology, potentially enabling
precise measurements of the dark energy parameters with a minimum of
systematic errors (Cooray et al.\ 2001; Eisenstein 2002; Blake \&
Glazebrook 2003; Seo \& Eisenstein 2003; Linder 2003).  The
large-scale linear clustering pattern contains a series of
small-amplitude, roughly sinusoidal, modulations in power of identical
physical origin to the acoustic peaks observed in the Cosmic Microwave
Background (CMB) (see Eisenstein \& Hu 1998; Meiksin, White \& Peacock
1999 and references therein).  These oscillations encode a
characteristic scale -- the sound horizon at recombination -- which
can be accurately calibrated using the linear physics of the CMB.  The
apparent value of this preferred scale, deduced from a slice of a
galaxy spectroscopic redshift survey, depends on the assumed
cosmological distance and expansion rate at the slice redshift $z$,
which control the mapping of redshifts to physical co-ordinates in the
tangential and radial directions, respectively.  The baryon
oscillations can therefore be used to measure the angular diameter
distance $D_A(z)$ and Hubble parameter $H(z)$ in units of the sound
horizon, over a series of redshift slices.  The acoustic signature may
also be measured from photometric redshift surveys (Blake \& Bridle
2005), although the smearing of radial information implies that only
$D_A(z)$ may be determined with any confidence.  The preferred scale
was recently identified in the clustering pattern of Luminous Red
Galaxies in the Sloan Digital Sky Survey (Eisenstein et al.\ 2005) and
used to constrain cosmological parameters for the first time.  A power
spectrum analysis of the final 2-degree Field Galaxy Redshift Survey
produced consistent results (Cole et al.\ 2005).

A number of techniques have been employed to estimate the accuracy
with which the baryon oscillation scales may be determined by future
galaxy surveys.  Blake \& Glazebrook (2003, hereafter BG03) used a
Monte Carlo, semi-empirical approach in which realizations of
spectroscopic redshift surveys were created from an underlying linear
power spectrum.  The acoustic scale was recovered for each realization
by first dividing out the overall shape of the measured power
spectrum, then fitting a simple empirically-motivated decaying
sinusoid to the baryon oscillations, up to a maximum wavenumber
determined by a conservative estimate of the extent of the linear
regime at the redshift in question.  The scatter in the best-fitting
values of the acoustic `wavelength' across the realizations represents
the accuracy with which the preferred scale may be extracted in such
an experiment.  Blake \& Bridle (2005) extended this methodology to
photometric redshift surveys.

A feature of this `model-independent' approach is that the information
contained in the oscillations is decoupled from that encoded by the
overall shape of the power spectrum, which is divided out prior to
fitting the sinusoid, and which may be subject to smooth broad-band
systematic tilts from such effects as poorly-modelled redshift-space
distortions, scale-dependent bias and non-linear growth of structure.
On the other hand, the power spectrum shape does also depend on the
cosmological parameters, and combined measurements in the tangential
and radial directions permit an Alcock-Paczynski test (Yamamoto,
Bassett \& Nishioka 2005).  BG03 discard this potentially useful
information for the benefit of simulated measurements that are more
`robust' against the presence of systematic errors.  Furthermore the
BG03 analysis contains various approximations, as discussed below.

Various other studies predicting the standard ruler accuracies from
baryon oscillation surveys have fitted a full power spectrum template
in order to estimate the cosmic distance scale.  Several papers (Seo
\& Eisenstein 2003; Hu \& Haiman 2004; Amendola, Quercellini \&
Giallongo 2005; Huetsi 2005) employ Fisher matrix techniques to
recover predicted errors in the cosmological quantities.  In addition,
galaxy catalogues extracted from N-body simulations, and therefore
incorporating realistic non-linear and galaxy biasing effects, have
been analyzed by Angulo et al.\ (2005), Springel et al.\ (2005), Seo
\& Eisenstein (2005) and White (2005).  In these studies the power
spectrum is typically `linearized' using a polynomial function, and is
then fitted with a linear-regime model using normal chi-squared
techniques.  Whilst the initial results from these investigations are
encouraging, it appears that redshift-space distortions slightly
degrade the baryon oscillation accuracy in the radial direction when
fitting a full power spectrum template (Seo \& Eisenstein 2005).
Larger simulations are required in order to quantify accurately the
potential influence of low-level systematic errors from redshift-space
distortions, scale-dependent bias and non-linear growth of structure
in the baryon oscillations technique.

The `model-independent' and `full template' methods are complementary,
with the minimal sinusoid-fitting providing an effective lower limit
to the efficacy of the technique, and the full power spectrum shape
fit indicating what may be achieved with more assumptions.  In fact,
in the regime where the oscillations are measured with high
statistical confidence, they encode most of the potential for
constraining cosmology (Hu \& Haiman 2004) and the accuracies
predicted by the two techniques agree reasonably well (Glazebrook \&
Blake 2005).

The simplicity of the `model-independent' technique of BG03 implies
that the resulting baryon oscillation accuracies scale in a relatively
predictable manner with the spectroscopic and photometric survey
parameters: central redshift, survey volume, galaxy number density and
(where applicable) photometric redshift error.  The purpose of this
study is to provide accurate fitting formulae for these standard ruler
accuracies in terms of the survey parameters.  We consider baryon
oscillation measurements in both the tangential and radial directions
for spectroscopic redshift surveys, and in the tangential direction
alone for photometric redshift surveys.  These formulae will be
considerably more efficient to implement than full Monte Carlo power
spectrum realizations, and should prove useful for planning future
galaxy surveys with the goal of measuring the dark energy parameters.
The general design optimization of such surveys involves consideration
of a potentially vast parameter space of survey configurations and
cosmological models (see Bassett 2005; Bassett, Parkinson \& Nichol
2005), and in this context a fitting formula is invaluable.

We assume a fiducial $\Lambda$CDM flat cosmological model with matter
density $\Omega_{\rm m} = 0.3$ although, as discussed in Section
\ref{seccosmo}, our results apply more generally.

\section{Simulated galaxy surveys}
\label{secsurv}

\subsection{Starting assumptions}

We begin by emphasizing some approximations inherent in our analysis:

\begin{itemize}

\item We utilize no information encoded in the overall shape of the
  power spectrum, which is divided out by a smooth polynomial prior to
  the baryon oscillation fit.

\item We assume that the power spectrum errors can be described by
  (correlated) Gaussian statistics for wavenumbers $k$ up to a maximum
  $k_{\rm max}$, specified by a conservative estimate of the extent of
  the linear regime at the redshift in question, and that modes with
  scales $k > k_{\rm max}$ provide no information.

\item We employ an approximate parameterized fit (a decaying sinusoid)
  for the baryon oscillation signature (see BG03).  We neglect the
  small scale-dependent phase shifts of the acoustic peaks and
  troughs.

\item We assume that shot noise can be described by Poisson statistics
  at the per cent level.  This may not be the case, and planned galaxy
  surveys may require slightly higher galaxy number densities in order
  to be cosmic-variance limited.

\end{itemize}

The measurements of future galaxy surveys should be fitted with
accurate templates, marginalizing over model uncertainities, rather
than empirical sinusoids.  However, we believe that the approximations
contained in our current analysis are acceptable because:

\begin{itemize}

\item Comparing our results with full Fisher matrix simulations, such
  as those of Seo \& Eisenstein (2003, 2005), our inferred standard
  ruler accuracies are comparable (albeit $30 - 50\%$ larger,
  reflecting our more conservative approach).

\item A major simulation effort is still required to model halo bias,
  non-linear structure formation and redshift-space distortions to the
  required accuracy (together with their effects on power spectrum
  mode correlations).  Using this information in our templates at this
  stage would create the possibility of additional systematic error.

\item Our approach enables us to explore a very large number of survey
  configurations and cosmological models.

\end{itemize}

\subsection{The simulation grid}

In order to explore the scalings of baryon oscillation accuracies with
galaxy survey parameters, we created a large grid of simulated
spectroscopic and photometric redshift surveys.  For each survey
configuration we used Monte Carlo realizations to determine the
accuracy with which the standard ruler could be measured, as described
below.  For spectroscopic surveys, we derived accuracies in the
tangential and radial directions.  For photometric surveys, the
damping of radial information implies that baryon oscillations may
only be measured in the tangential direction (Seo \& Eisenstein 2003,
Blake \& Bridle 2005).

In cosmological terms, the measurement accuracies of the tangential
and radial baryon oscillation scales determine the precision with
which the quantities $r(z)/s$ and $r'(z)/s$ may be inferred, where
$r(z)$ is the co-moving distance to the redshift slice $z$, $r'(z)
\equiv dr/dz = c/H(z)$ where $c$ is the speed of light and $H(z)$ is
the Hubble parameter measured by an observer at redshift $z$, and $s$
is the (co-moving) sound horizon at recombination.  Given that $s =
\theta_{\rm A} \times r({\rm CMB})$, where $\theta_{\rm A}$ is the
(accurately-known) angular scale of the first CMB acoustic peak and
$r{(\rm CMB)}$ is the inferred distance to the surface of last
scattering, then the acoustic oscillations may be thought of as
measuring the quantities $r(z)/r({\rm CMB})$ and $r'(z)/r({\rm CMB})$
(see Eisenstein et al.\ 2005).

Our grid of Monte Carlo simulations was generated by varying four
survey parameters: the central redshift $z$, the survey area $A$ (in
$10^3$ deg$^2$), the survey width $\delta z$ (such that the survey
ranges between redshifts $z - \delta z$ and $z + \delta z$) and the
number density of galaxies $n$ (in $10^{-3} \, h^3$ Mpc$^{-3}$).  For
photometric redshift surveys, a fifth parameter was added: the
r.m.s.\ error in redshift $\sigma_z$, expressed via the parameter
$\sigma_0 = \sigma_z/(1+z)$.  Given the potential complexity of target
selection techniques for these surveys, we do not consider realistic
galaxy redshift distributions, but instead populate the survey volume
uniformly with number density $n$.  This will be a good approximation
for any relatively narrow survey redshift slice (or for high-enough
number density the measurements will be cosmic-variance limited and
independent of $n$).  Further details of our simulations of
spectroscopic and photometric redshift surveys are given below.

\subsection{Spectroscopic redshift surveys}
\label{secsurvspec}

The trial parameter values for the simulated spectroscopic redshift
surveys were:
\begin{eqnarray}
z &=& 0.2 \rightarrow 3.4 \hspace{0.3cm} {\rm in \: steps \: of \:
  0.2} \nonumber \\ A &=& 0.1, 0.25, 0.5, 0.75, 1.0, 1.5, 2.0, 3.0,
4.0, 5.0 \nonumber \\ \delta z &=& 0.025, 0.05, 0.075, 0.1, 0.125,
0.15, 0.2, 0.3, 0.4, 0.5 \nonumber \\ n &=& 0.04, 0.08, 0.16, 0.32,
0.64, 1.3, 2.6, 5.1, 10.2, 20.6 \nonumber
\end{eqnarray}
i.e., $16{,}600$ configurations were analyzed (given that cases with
$\delta z > z$ are excluded).

The Monte Carlo methods utilized to analyze the simulated
spectroscopic redshift surveys were similar to those employed by BG03
and described in more detail by Glazebrook \& Blake (2005).  However,
some improvements in speed were required to process such a large grid
of surveys.  In BG03, Monte Carlo realizations of surveys were
generated by performing Gaussian realizations of the underlying power
spectrum, and Poisson sampling the resulting density fields.  The
covariance matrix of the power spectrum bins was effectively
determined numerically by averaging over the Monte Carlo realizations.
This process is too time-consuming for exploring a large grid of
surveys.  Therefore in this study we determined the covariance matrix
analytically from the Fourier transform of the survey window function,
by evaluating the sums given in Feldman, Kaiser \& Peacock (1994), who
present an optimal estimator for the power spectrum.  Our analysis
assumes a full conical survey geometry, and hence includes the
convolution of the underlying power spectrum with the survey geometry
and the correlations between the Fourier bins.

In more detail: for a given survey configuration we enclosed the
survey cone (sampled by a uniform number density $n$) with a cuboid of
volume $V_{\rm cub}$, and determined the Fourier transform of the
window function, which we write as $W_k$.  We assumed a model linear
power spectrum given by the fitting formula of Eisenstein \& Hu
(1998), using fiducial cosmological parameters $\Omega_{\rm m} = 0.3$,
$\Omega_\Lambda = 0.7$, $h = 0.7$, $\Omega_{\rm b}/\Omega_{\rm m} =
0.15$, $n_{\rm s} = 1$ and $\sigma_8 = 1$.  We scaled this power
spectrum to redshift $z$ using the linear growth factor $D(z)$ of
Carroll, Press \& Turner (1992).  We convolved the input power
spectrum with the survey window function.  We restricted our analysis
to scales larger than a maximum wavenumber $k$ corresponding to a
conservative estimate of the transition scale between the linear and
non-linear clustering regimes (see BG03).  Our surveys assume a linear
bias factor $b_0 = 1$ for galaxies with respect to matter, which is
likely to be conservative at high redshifts.  Our fitting formulae can
be simply adapted for $b_0 \ne 1$ as explained below.

In order to obtain the power spectrum covariance matrix, we note that
equation $2.5.2$ in Feldman, Kaiser \& Peacock (1994) reduces to the
expression (see also Tadros \& Efstathiou 1996):
\begin{equation}
C_{ij} \equiv \frac{<\delta P_i \, \delta P_j>}{(P + \frac{1}{n})^2} =
\frac{2}{N_{\rm sum}} \frac{\sum_{k,k'} |W_{k - k'}|^2}{\sum_k
  |W_k|^2}
\label{eqpkcov}
\end{equation}
where $P \equiv P(k)$ is the power spectrum amplitude typical of bins
$i$ and $j$, the summation in the numerator is evaluated between all
$N_{\rm sum}$ separate pairs of modes $(k,k')$ in bins $i$ and $j$,
and the summation in the denominator is evaluated over all Fourier
modes and is equal to $(V/V_{\rm cub})^2$ for a window function which
is either a uniform value or zero.  The factor of 2 reflects the fact
that only half of the measured Fourier modes are independent, owing to
the reality condition of the density field.  For a survey in a uniform
box, $W_k = 0$ unless $k=0$ and equation \ref{eqpkcov} reduces to
$C_{ii} = 2/m$, where $m$ is the total number of Fourier modes in bin
$i$.  We evaluated the double sum in equation \ref{eqpkcov} using a
Monte Carlo integration scheme.  We binned the power spectrum into a
2-dimensional grid of radial and tangential components (if the
$x$-axis is the radial direction, then $k_{\rm rad} = |k_x|$ and
$k_{\rm tan} = \sqrt{k_y^2 + k_z^2}$).  We used Fourier bin widths
$\Delta k_{\rm rad} = \Delta k_{\rm tan} = 0.01 \, h$ Mpc$^{-1}$,
unless a survey dimension $L$ was sufficiently small that the
corresponding spacing of the Fourier modes $2\pi/L > 0.005 \, h$
Mpc$^{-1}$, in which case we set $\Delta k = 4\pi/L$ (i.e., the
minimum thickness of a bin in our analysis is 2 Fourier modes).  Given
that the acoustic `wavelength' in Fourier space is $k_{\rm A} \approx
0.06 \, h$ Mpc$^{-1}$, we rejected a survey configuration if $\Delta k
> 0.03 \, h$ Mpc$^{-1}$.  We tested our code using an
analytically-tractable survey window function for which the sums in
equation \ref{eqpkcov} could be evaluated in closed form.

Having determined the covariance matrix, we created many Gaussian
realizations of correlated power spectrum measurements using the
technique of Cholesky decomposition.  The acoustic `wavelengths' in
the tangential and radial directions were fit to these realizations
via a simple empirical decaying sinusoid, using the same method as
Glazebrook \& Blake (2005), and the accuracies with which these scales
could be measured was inferred using the scatter in the resulting
best-fitting `wavelengths' across the realizations.  Tests repeating
the analysis for identical survey configurations showed that the
scatter in determination of the standard ruler accuracy $y$ owing to
numerical noise, due to the approximate summation scheme for equation
\ref{eqpkcov} and the finite number (400) of power spectrum
realizations, was about $5\%$ of $y$ ($7\%$ of $y$) in the tangential
(radial) direction.

\subsection{Photometric redshift surveys}

The trial parameter values for the simulated photometric redshift
surveys were:
\begin{eqnarray}
z &=& 0.5, 0.75, 1.0, 1.25, 1.5, 1.75, 2.0, 2.5, 3.0 \nonumber \\ A
&=& 0.5, 0.75, 1.0, 2.0, 5.0, 10.0 \nonumber \\ \delta z &=& 0.1, 0.2,
0.3, 0.4, 0.5 \nonumber \\ n &=& 0.1, 0.2, 0.5, 1.0, 2.0, 5.0, 10.0
\nonumber \\ \sigma_0 &=& 0.01, 0.02, 0.03, 0.04, 0.05 \nonumber
\end{eqnarray}
i.e., $9{,}450$ configurations were analyzed.

For this set of Monte Carlo simulations a flat-sky approximation was
assumed, as described by Blake \& Bridle (2005).  Therefore, unlike
for the spectroscopic survey analysis, no window function effects are
considered.  The photometric redshift error distribution was assumed
to be a Gaussian function, hence the power spectrum was assumed to be
damped in the radial direction by a factor $\exp{[-(k_{\rm rad}
    \sigma_r)^2]}$, where $\sigma_r = \sigma_0 (1+z) \times dr/dz$
(Blake \& Bridle 2005, equations 3 and 4).

In order to speed up the computation in the same style as described
above for spectroscopic surveys, the error in the power spectrum
measurement for photometric redshift surveys was not determined via
radially-smeared Monte Carlo realizations of the density field (as in
Blake \& Bridle 2005) but instead using an analytical approximation.
The error in the power spectrum measured in a Fourier cell centred at
$(k_{\rm tan},k_{\rm rad})$ of width $(\delta k_{\rm tan}, \delta
k_{\rm rad})$ was assumed to be the usual combination of cosmic
variance and shot noise:
\begin{equation}
\delta P = \frac{1}{\sqrt{m}} \left( P \exp{[-(k_{\rm rad}
    \sigma_r)^2]} + \frac{1}{n} \right)
\label{eqpkerr}
\end{equation}
where $P \equiv P(k) = P(\sqrt{k_{\rm tan}^2 + k_{\rm rad}^2})$ is the
(undamped) value of the model power spectrum in the bin, taken from
the fitting formula of Eisenstein \& Hu (1998), and $m$ is the number
of Fourier modes contributing (i.e.\ those contained in an annulus of
radius $k_{\rm tan}$, radial thickness $\delta k_{\rm tan}$ and length
$\delta k_{\rm rad}$).  The value of $m$ was determined from the
density-of-states in $k$-space, $\rho_k = V/(2\pi)^3$: $m = \rho_k
\times 2 \pi k_{\rm tan} \times \delta k_{\rm tan} \times \delta
k_{\rm rad}$.

For each survey configuration, we generated a large number of Monte
Carlo realizations of noisy power spectra by adding a Gaussian
variable of standard deviation $\delta P$ (given by equation
\ref{eqpkerr}) to the damped model power spectrum $P \exp{[-(k_{\rm
      rad} \sigma_r)^2]}$.  We binned each power spectrum realization
in tangential Fourier bins of width $\delta k_{\rm tan} = 0.01 \, h$
Mpc$^{-1}$ by averaging cells in the radial direction up to a maximum
value of $k_{\rm rad} = 1.5/\sigma_r$, beyond which the power spectrum
contains very little signal owing to the damping.  For each
realization, we then divided the binned power spectrum by a smooth
`reference spectrum' and fitted the result with the empirical decaying
sinusoid used in BG03.  As in the case of the spectroscopic redshift
surveys, the scatter in the best-fitting values of the sinusoidal
`wavelength' across the Monte Carlo realizations was taken as the
tangential baryon oscillation accuracy for this survey configuration.
Tests repeating the analysis for identical configurations showed that
the scatter in the determination of the standard ruler accuracy $y$
owing to numerical noise was about $5\%$ of $y$.

\section{The fitting formula}
\label{secexpress}

\begin{table*}
\center
\caption{Best-fitting coefficients for the fitting formula, defined by
  equations \ref{eqfitform3} to \ref{eqfitform6}, for the three types
  of standard ruler accuracy.  The first set of parameters, from $x_0$
  to $b$, apply only to high-accuracy baryon oscillation measurements
  where the simple scalings described by equation \ref{eqfitform3} are
  valid (i.e., $x \ll x_t$).  The second set of parameters, from $p$
  to $\beta$, describe the worsening standard ruler performance in the
  regime where the oscillations are just being resolved (see equation
  \ref{eqfitform5}).  The r.m.s.\ difference in the predictions of the
  fitting formulae and Monte Carlo simulations is listed separately
  for measurements of all accuracies $(y < 10\%)$ and of just high
  accuracy ($y < 2\%$).}
\label{tabparam}
\begin{tabular}{cccc}
\hline
Parameter & Spec-$z$ & Spec-$z$ & Photo-$z$ \\
& Tangential & Radial & Tangential \\
\hline
$x_0$ (per cent) & 0.85 & 1.48 & 1.23 \\
$n_0 (\times 10^{-3} \, h^3 \, {\rm Mpc}^{-3})$ & 0.82 & 0.82 & 0.71 \\
$z_{\rm m}$ & 1.4 & 1.4 & 1.4 \\
$\gamma$ & 0.5 & 0.5 & 0.61 \\
$b$ & 0.52 & 0.52 & 0.52 \\
\hline
$p$ & 2 & 2 & 4 \\
$a$ & 7.3 & 10.6 & 4.2 \\
$\alpha$ & 0.26 & 0.49 & 0.11 \\
$\beta$ & 0.27 & 1.00 & 0.42 \\
\hline
r.m.s.\ error in fitting formula ($y < 10\%$) & 7.1\% of $y$ & 9.9\% of $y$ & 6.9\% of $y$ \\
r.m.s.\ error in fitting formula ($y < 2\%$) & 5.5\% of $y$ & 7.4\% of $y$ & 4.0\% of $y$ \\
\hline
\end{tabular}
\end{table*}

In this Section we develop an analytic expression for the accuracy of
measurement of the baryon oscillation scale in terms of the survey
configuration: central redshift $z$, total volume $V$ (in $h^{-3}$
Gpc$^3$), average number density of galaxies $n$ (in $10^{-3} \, h^3$
Mpc$^{-3}$) and, in the case of photometric redshift surveys, the
r.m.s.\ error in co-moving co-ordinate $\sigma_r$ (in $h^{-1}$ Mpc).
The fitting formula contains free parameters whose values are
calibrated using the grids of simulated surveys described in Section
\ref{secsurv}.  Different fitting formula coefficients were derived
for standard ruler accuracies resulting from spectroscopic redshift
surveys (separately for the tangential and radial directions) and
photometric redshift surveys (in the tangential direction only).

As a first approximation for the fitting formula, we assumed that the
accuracy $x$ with which the acoustic scale can be measured is
proportional to the average fractional error $\delta P/P$ in the power
spectrum, given by the usual sampling formula (e.g.\ Tegmark 1997):
\begin{equation}
x \propto \frac{\delta P}{P} = \frac{1}{\sqrt{m}} \left( 1 +
\frac{1}{nP} \right)
\label{eqfitform1}
\end{equation}
where $m$ is the total number of independent Fourier modes
contributing to the measurement, and $P \equiv P(k^*)$ (in $h^{-3}$
Mpc$^3$) is the value of the power spectrum amplitude at an average
scale $k^* \simeq 0.2 \, h$ Mpc$^{-1}$ characteristic of the baryon
oscillations.  The two terms in equation \ref{eqfitform1} represent
the effects of cosmic variance and shot noise, respectively.  Given a
fixed amount of observing time, optimal measurements of the power
spectrum follow from a survey of depth such that $nP \sim 1$.  This
requirement is readily achieved by $\sim 1$ hr integrations with 8-m
class ground-based telescopes.

The number of measured Fourier modes scales with the total survey
volume $V$, which determines the density-of-states in Fourier space
$\rho_k$: $m \propto \rho_k \propto V$.  In addition, $m$ is
proportional to the contributing volume in $k$-space.  For a fiducial
survey spanning $1000$ deg$^2$ from $z = 0.6$ to $z = 1.4$
(i.e.\ $A=1$, $\delta z = 0.4$), we find $V = 2.16 \, h^{-3}$ Gpc$^3$
for our fiducial cosmological parameters.  The number of independent
modes contained in an angle-averaged Fourier bin at $k = 0.2 \, h$
Mpc$^{-1}$ of width $\delta k = 0.01 \, h$ Mpc$^{-1}$ is then $m
\approx 2.2 \times 10^4$, yielding a measurement of the power spectrum
in this bin with an accuracy of about $1\%$ using equation
\ref{eqfitform1}.  As the baryon oscillations have a fractional
amplitude of roughly $5\%$, this constitutes a high-significance
detection.

For photometric redshift surveys, the radial smearing damps out the
useful signal in all Fourier modes with small-scale radial $k$-values
$k_{\rm rad} \ga 1/\sigma_r$.  Therefore, $m \propto 1/\sigma_r$.  For
a typical photometric redshift performance, $1/\sigma_r \sim 0.02 \,
h$ Mpc$^{-1}$.  Comparing this scale with the extent of the available
linear regime, $k \la 0.2 \, h$ Mpc$^{-1}$, we find that the number of
usable Fourier modes is diminished by roughly an order of magnitude
for photometric surveys.

If we also include the scaling of the power spectrum with redshift as
the linear growth factor $D(z)$ and a linear bias parameter $b_0$, we
can re-write equation \ref{eqfitform1} as:
\begin{equation}
x \propto \frac{\sqrt{\sigma_r}}{\sqrt{V}} \left( 1 + \frac{n_{\rm
    eff}}{n} \frac{D(z_0)^2}{b_0^2 D(z)^2} \right)
\label{eqfitform2}
\end{equation}
where $n_{\rm eff}$ ($\sim 1/P(k^*)$ at $z = z_0$) is a fiducial
number density (a fitted parameter) and $D(z_0)$ is the linear growth
factor at a fiducial redshift $z = z_0$.  The factor $\sqrt{\sigma_r}$
only appears for photometric redshift surveys.  The assumption of a
linear scale-independent bias factor $b_0$ will be incorrect in
detail.  However, given that in our simulations the overall shape of
the power spectrum is divided out prior to fitting the sinusoidal
function, our results are not sensitive to such details.  The value of
$b_0$ may be interpreted as the boost in the power spectrum of
galaxies with respect to that of dark matter at the characteristic
scale of the baryon oscillations, $k^* \simeq 0.2 \, h$ Mpc$^{-1}$:
$b_0 = \sqrt{P_{\rm gal}(k^*)/P_{\rm dm}(k^*)}$.  We note that the
quantity $b_0 \times D(z)$ in equation \ref{eqfitform2} is observed to
be roughly constant with redshift for $L^*$ galaxies (Lahav et
al.\ 2002).

We normalize equation \ref{eqfitform2} to an accuracy $x = x_0$ for a
fiducial survey of volume $V = V_0$ and (where applicable) redshift
error $\sigma_r = \sigma_{r,0}$:
\begin{equation}
x = x_0 \sqrt{\frac{V_0}{V}} \sqrt{\frac{\sigma_r}{\sigma_{r,0}}}
\left( 1 + \frac{n_{\rm eff}}{n} \frac{D(z_0)^2}{b_0^2 D(z)^2} \right)
\end{equation}
We take $z_0 = 1$ and the same fiducial survey as above ($A=1$,
$\delta z = 0.4$) such that $V_0 = 2.16 \, h^{-3}$ Gpc$^3$ and $D(z_0)
= 0.61$ (Carroll, Press \& Turner 1992).  We also assume $\sigma_{r,0}
= 34.1 \, h^{-1}$ Mpc, corresponding to a redshift error $\sigma_z =
0.01 (1+z)$ at $z = z_0$.

The fitting formula must also reflect the increase with redshift of
the extent of the linear regime (i.e.\ the number of acoustic peaks
which may be fitted), which enables a more accurate determination of
the acoustic scale for a fixed power spectrum precision (see
Glazebrook \& Blake 2005, Figure 7).  This is accomplished via an
empirical power-law in $z$, which is cut off at a maximum redshift $z
= z_{\rm m}$ at which all of the high-amplitude peaks are visible:
\begin{eqnarray}
x \hspace{-0.2cm} &=& \hspace{-0.2cm} x_0 \sqrt{\frac{V_0}{V}}
\sqrt{\frac{\sigma_r}{\sigma_{r,0}}} \left( 1 + \frac{n_{\rm eff}}{n}
\frac{D(z_0)^2}{b_0^2 D(z)^2} \right) \left( \frac{z_{\rm m}}{z}
\right)^\gamma \hspace{0.1cm} z < z_{\rm m} \nonumber
\\ &=& \hspace{-0.2cm} x_0 \sqrt{\frac{V_0}{V}}
\sqrt{\frac{\sigma_r}{\sigma_{r,0}}} \left( 1 + \frac{n_{\rm eff}}{n}
\frac{D(z_0)^2}{b_0^2 D(z)^2} \right) \hspace{1.2cm} z > z_{\rm m}
\label{eqfitform3}
\end{eqnarray}
where $\gamma > 0$ is a fitted parameter.  In addition, given that the
amplitude of the power spectrum decreases with increasing $k$, the
variation in the extent of the linear regime with $z$ changes the
average amplitude of $P(k)$ included in the analysis, and hence the
value of $n_{\rm eff}$.  We described this variation by
\begin{eqnarray}
n_{\rm eff} \hspace{-0.2cm} &=& \hspace{-0.2cm} n_0 \left[ 1 - b
  \left( 1 - \frac{z}{z_{\rm m}} \right) \right] \hspace{0.5cm} z <
z_{\rm m} \nonumber \\ &=& \hspace{-0.2cm} n_0 \hspace{3.15cm} z >
z_{\rm m}
\label{eqfitform4}
\end{eqnarray}
The fitting formulae of equations \ref{eqfitform3} and
\ref{eqfitform4}, containing five free parameters ($x_0$, $n_0$, $b$,
$\gamma$, $z_{\rm m}$), work well for high-precision measurements of
the acoustic scale.  However, in the regime where the oscillations are
just being resolved, the scaling of the `model-independent' accuracy
with (for example) survey volume is more rapid than $V^{-1/2}$.  We
therefore modified the accuracy $x$ to a new value $y$ where
\begin{equation}
y = \frac{x}{1 - \left( \frac{x}{x_t} \right)^p}
\label{eqfitform5}
\end{equation}
where $x_t$ is a characteristic accuracy and $p > 0$ is a free
parameter.  Equation \ref{eqfitform5} is designed such that $y
\rightarrow x$ as $x \rightarrow 0$.  The quantity $y$ is the final
predicted standard ruler accuracy of the fitting formula.

Empirically, we found that the quantity $x_t$ has a dependence on
survey volume $V$ and redshift $z$:
\begin{eqnarray}
x_t \hspace{-0.2cm} &=& \hspace{-0.2cm} a \left( \frac{V}{V_0}
\right)^\alpha \left( \frac{z_{\rm m}}{z} \right)^\beta \hspace{0.5cm}
z < z_{\rm m} \nonumber \\ &=& \hspace{-0.2cm} a \left( \frac{V}{V_0}
\right)^\alpha \hspace{1.6cm} z > z_{\rm m}
\label{eqfitform6}
\end{eqnarray}
Equations \ref{eqfitform5} and \ref{eqfitform6} hence introduce
another four parameters ($p$, $a$, $\alpha$ and $\beta$) which
describe the departure from the simple scaling in the high-accuracy
regime.

We emphasize that planned baryon oscillation surveys should aim to
reach the regime in which the acoustic features have been properly
resolved and the high-accuracy scalings of equation \ref{eqfitform3}
apply (i.e., $x \ll x_t$).  Otherwise, the detection of oscillations
will be of poor significance in many Monte Carlo realizations of the
planned survey.  The modifications represented by equations
\ref{eqfitform5} and \ref{eqfitform6} are included to ensure that
equation \ref{eqfitform3} is not applied in the regime where the
oscillations are poorly detected, which would result in
over-optimistic predictions of the standard ruler accuracies.
Equation \ref{eqfitform3} has some elements in common with the formula
suggested by Bernstein (2005, equation 42) for the tangential baryon
oscillation accuracy.

\section{Fitting formula coefficients}

We varied the free parameters of the fitting formulae to obtain the
best fit to the grid of Monte Carlo simulated surveys (in the sense of
the lowest standard deviation of the fractional variation).  If the
Monte Carlo accuracy of the acoustic scale was poorer than $10\%$ for
a grid point, then that survey configuration was assumed to provide no
measurement of the baryon oscillations and was ignored in the fitting
process.  Of the $16{,}600$ spectroscopic survey configurations,
$5{,}441$ ($4{,}576$) were included in the determination of the
fitting formula coefficients in the tangential (radial) direction.  Of
the $9{,}450$ photometric survey configurations, $3{,}765$ were
included.

Results are listed in Table \ref{tabparam} for the cases of
spectroscopic surveys (tangential and radial directions) and
photometric surveys (tangential direction).  Considering the whole
regime of standard ruler accuracies better than $10\%$ ($y < 0.1$),
the r.m.s.\ difference in the predictions of the fitting formulae and
Monte Carlo simulations is about $7\%$ of $y$ ($10\%$ of $y$) in the
tangential (radial) direction.  For high-precision measurements with
accuracies better than $2\%$ ($y \approx x < 0.02$) the formulae
perform significantly better: in this regime the fitting formulae
recover the baryon oscillation accuracies to better than $\pm 0.1\%$.
Given that the numerical noise in the grid of simulated surveys,
resulting from the Monte Carlo realizations, is approximately $5\%$ of
$y$ ($7\%$ of $y$) in the tangential (radial) directions, thus
constituting a significant fraction of the scatter, these fitting
formulae perform remarkably well.

The parameter $z_{\rm m}$, which is the redshift at which the
improvement in the baryon oscillation accuracy (for fixed survey
volume and number density) saturates, was constrained to have the same
value for all types of survey.  For the best-fitting value, $z_{\rm m}
= 1.4$, a conservative estimate of the extent of the linear regime
(see BG03) is $k_{\rm lin} = 0.25 \, h$ Mpc$^{-1}$, encompassing
essentially the whole range of high-amplitude acoustic peaks.  The
parameters $n_0$ and $b$ were also constrained to be equal in the
tangential and radial directions for spectroscopic surveys.  The
best-fitting value, $n_0 = 8.2 \times 10^{-4} \, h^3$ Mpc$^{-3}$,
corresponds to an effective power spectrum $P \sim 1/n_0 \approx 1200
\, h^{-3}$ Mpc$^3$.  This is very reasonable, given that for $z = z_0
= 1$ (see equation \ref{eqfitform2}) the amplitude of the power
spectrum is $P(k,z) = P(k,0) D(z)^2 \approx 1200 \, h^{-3}$ Mpc$^3$ at
$k = 0.19 \, h$ Mpc$^{-1}$.

Comparing the accuracies of measuring the tangential acoustic scale
with spectroscopic and photometric galaxy surveys we find that,
assuming an identical number density and redshift range, the
photometric survey (with r.m.s.\ error in radial co-ordinate
$\sigma_r$) must cover an area exceeding the spectroscopic survey by a
factor $\simeq 2.1 (\sigma_r / 34.1 \, h^{-1} \, {\rm Mpc})$ to
produce the same level of tangential accuracy.  Blake \& Bridle (2005)
present a wider range of comparisons.

\begin{figure}
\center
\epsfig{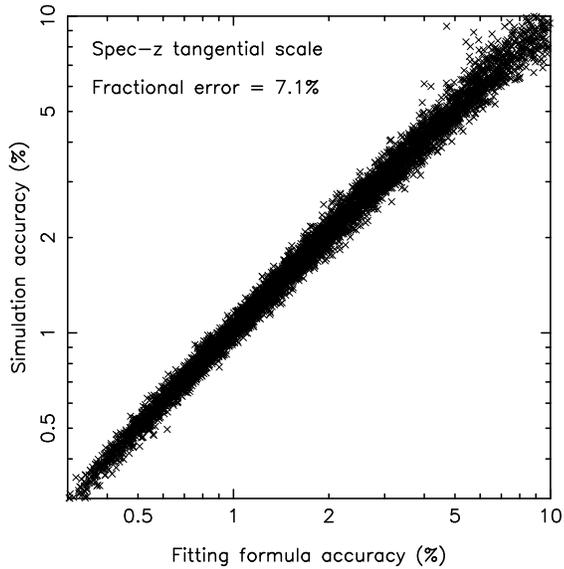}
\caption{Comparison of the fitting formula and Monte Carlo simulation
  accuracies of measuring the tangential acoustic scale from
  spectroscopic redshift surveys.  A significant fraction of the
  scatter is due to numerical noise in the simulations.}
\label{figfittan}
\end{figure}

\begin{figure}
\center
\epsfig{file=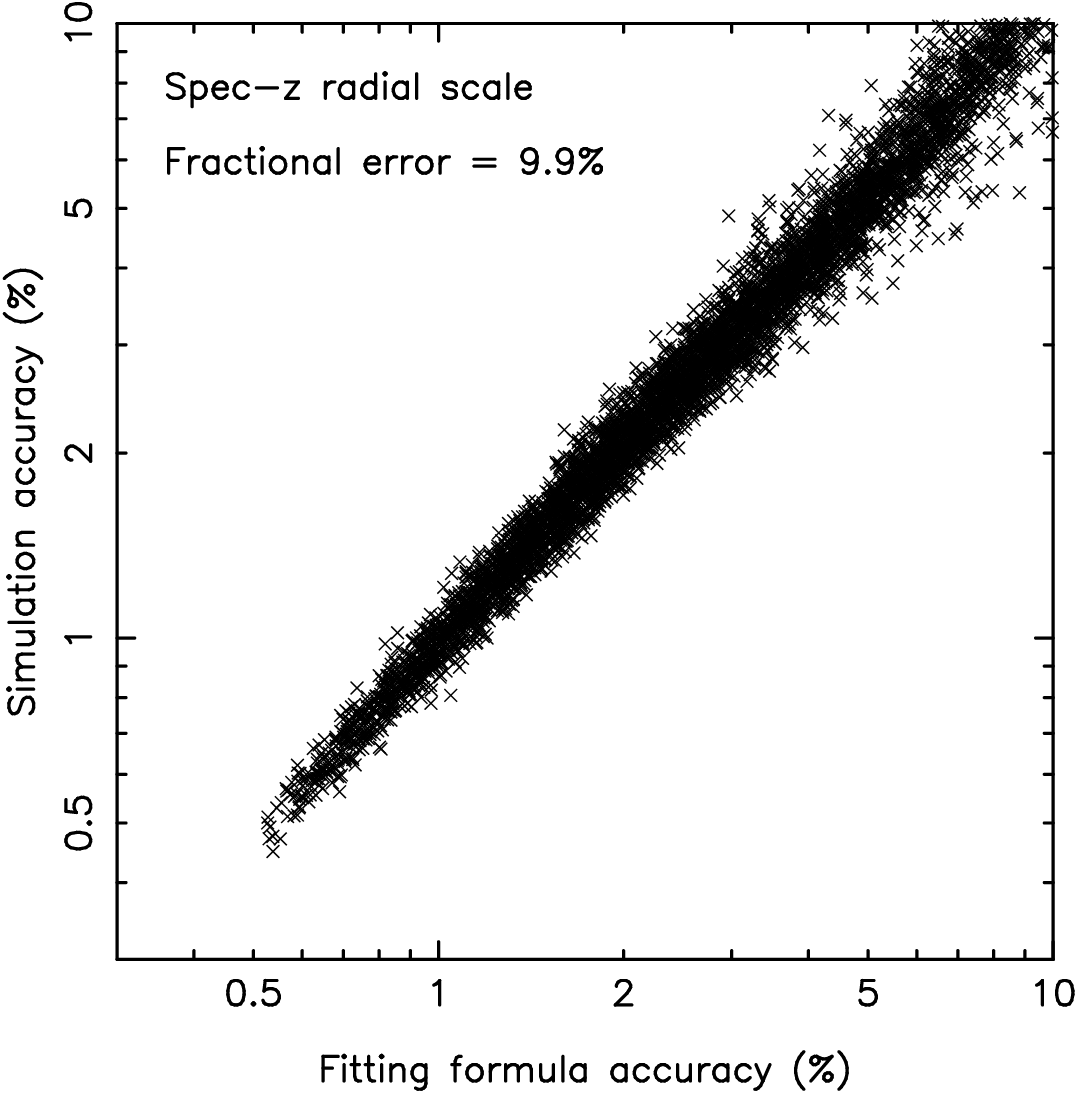,width=7.5cm,angle=-90}
\caption{Comparison of the fitting formula and Monte Carlo simulation
  accuracies of measuring the radial acoustic scale from spectroscopic
  redshift surveys.  A significant fraction of the scatter is due to
  numerical noise in the simulations.}
\label{figfitrad}
\end{figure}

\begin{figure}
\center
\epsfig{file=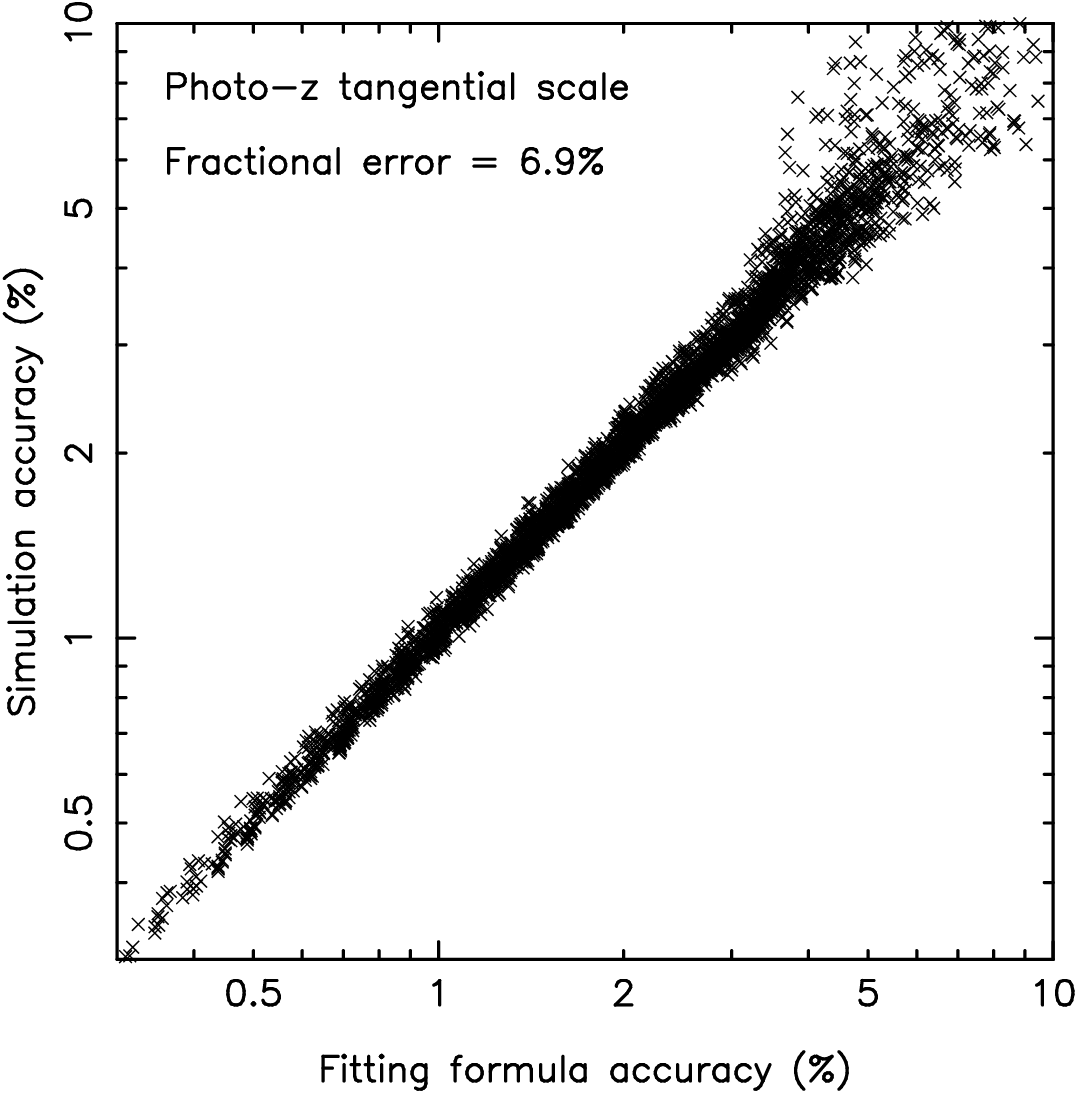,width=7.5cm,angle=-90}
\caption{Comparison of the fitting formula and Monte Carlo simulation
  accuracies of measuring the tangential acoustic scale from
  photometric redshift surveys.  A significant fraction of the scatter
  is due to numerical noise in the simulations.}
\label{figfitphot}
\end{figure}

\begin{figure}
\center
\epsfig{file=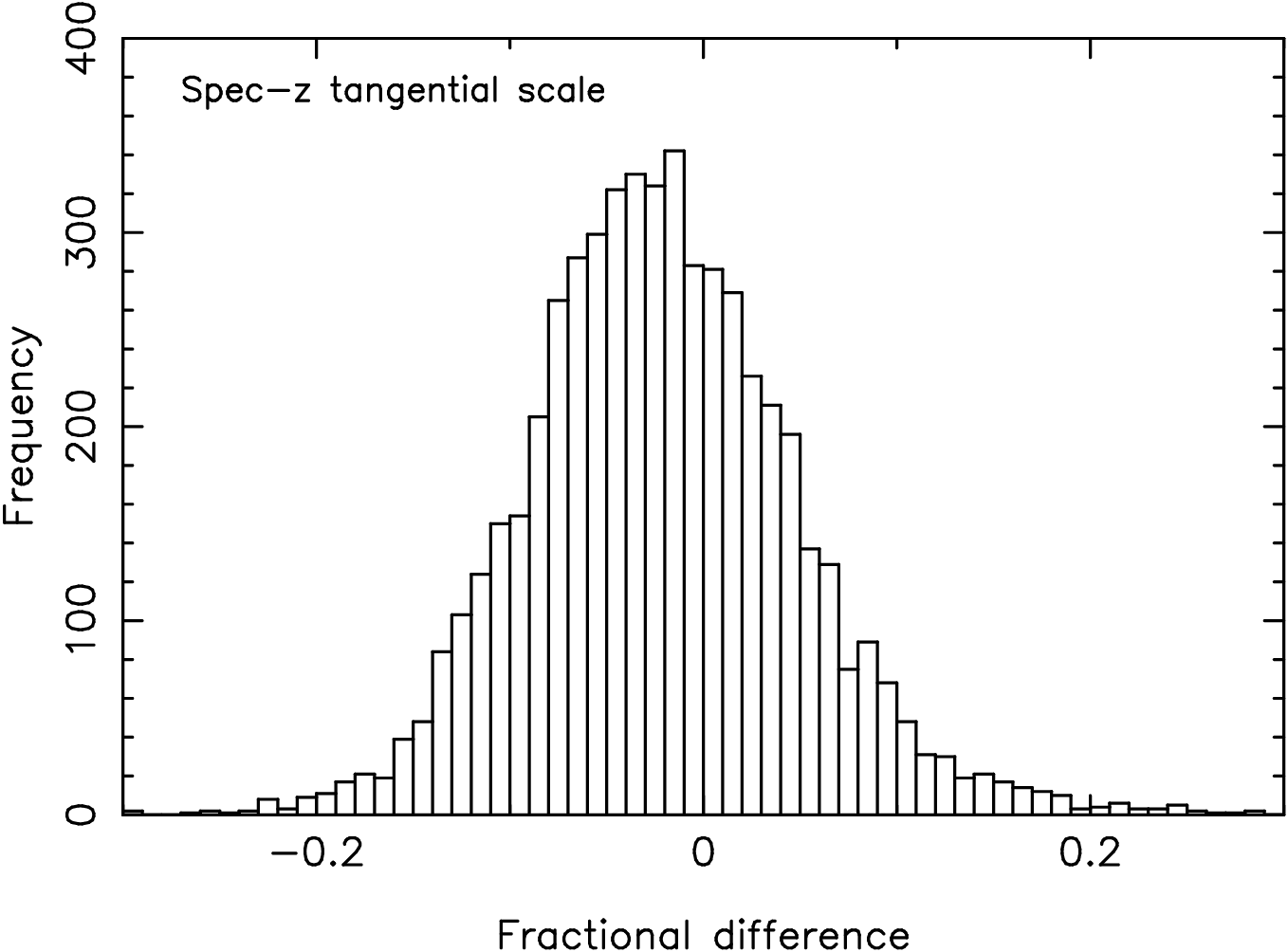,width=5.5cm,angle=-90}
\caption{Distribution of the fractional difference ($\Delta y/y$)
  between the fitting formula and Monte Carlo simulation accuracies of
  measuring the tangential acoustic scale from spectroscopic redshift
  surveys.  The scatter is well-described by a Gaussian distribution,
  with very few outliers.  The overall r.m.s.\ difference in
  accuracies is $7\%$ of $y$ (see Figure \ref{figfittan}).  There is a
  small offset in the mean difference ($2\%$ of $y$).}
\label{figfracdist}
\end{figure}

Figures \ref{figfittan} to \ref{figfitphot} compare the accuracies
predicted by the fitting formulae with those obtained from the grid of
simulated surveys for the three types of baryon oscillation
measurement, illustrating the tightness of the fits.  Figure
\ref{figfracdist} plots a histogram of the fractional difference in
the fitting formula and Monte Carlo accuracies for the case of the
tangential acoustic scale from spectroscopic surveys, demonstrating
that the scatter approximately follows a Gaussian distribution.  There
is a small systematic offset in the mean difference; when determining
the fitting formula coefficients we require that this offset is less
than $3\%$ of $y$.

\begin{figure}
\center
\epsfig{file=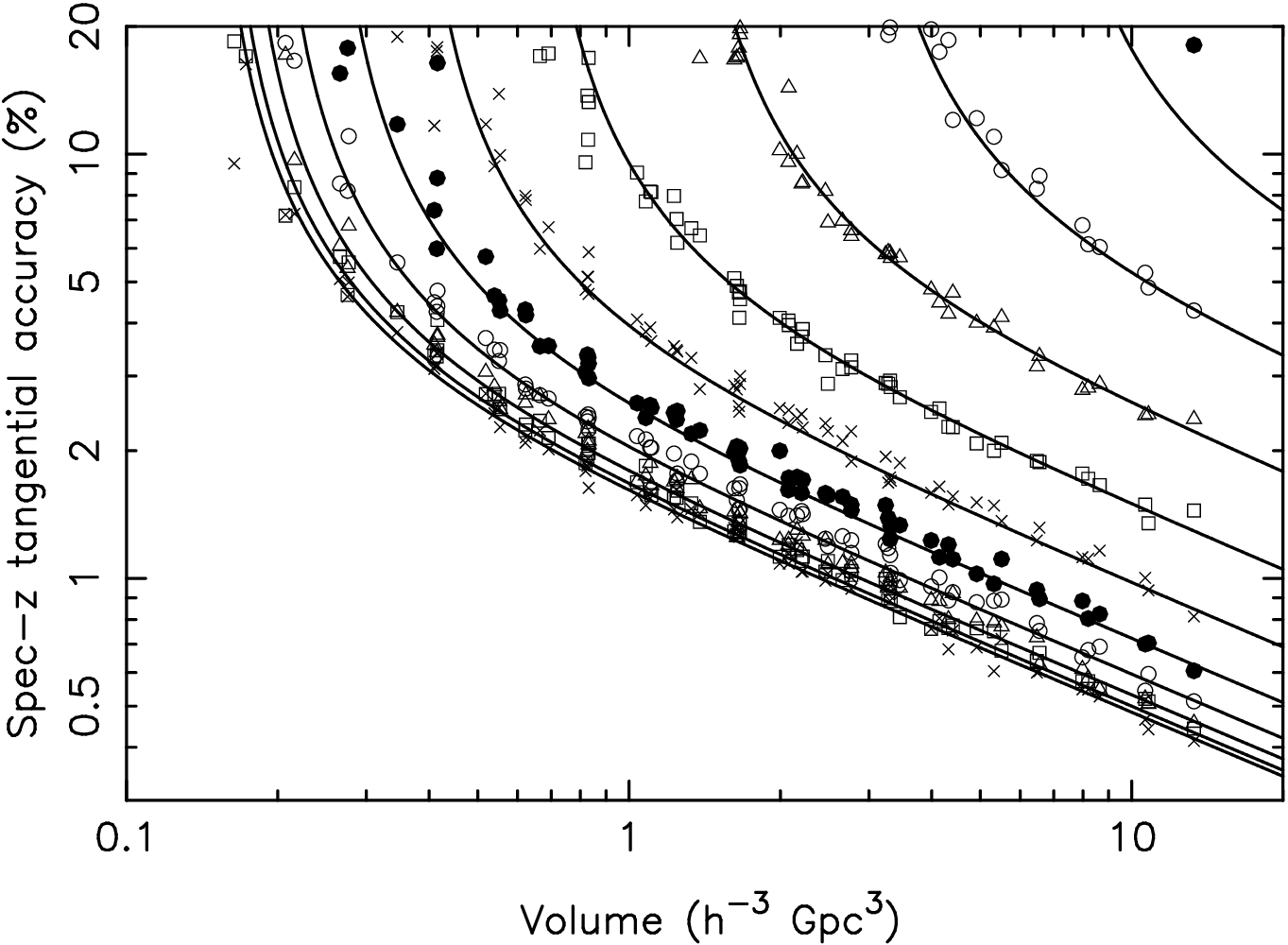,width=5.5cm,angle=-90}
\caption{Comparison of the fitting formula (lines) and Monte Carlo
  simulation accuracies (data points) of measuring the tangential
  acoustic scale from spectroscopic surveys at $z=1$, plotted as a
  function of survey volume.  The different lines and data point
  styles correspond to the 10 different values of the number density
  $n$ listed in Section \ref{secsurvspec}, ranging from low density
  (upper right) to high density (lower left).  The agreement between
  the fitting formula and Monte Carlo simulations is excellent.}
\label{fignpcurves}
\end{figure}

In order to demonstrate further the performance of the fitting
formulae, Figure \ref{fignpcurves} compares the predictions of the
formulae with the Monte Carlo data points for measurements of
tangential baryon oscillations from spectroscopic surveys at $z = 1$,
as a function of survey volume.  The various curves (and point styles)
correspond to different values of number density $n$.  The agreement
in the shape and offset of the curves is excellent.

\begin{figure*}
\center
\epsfig{file=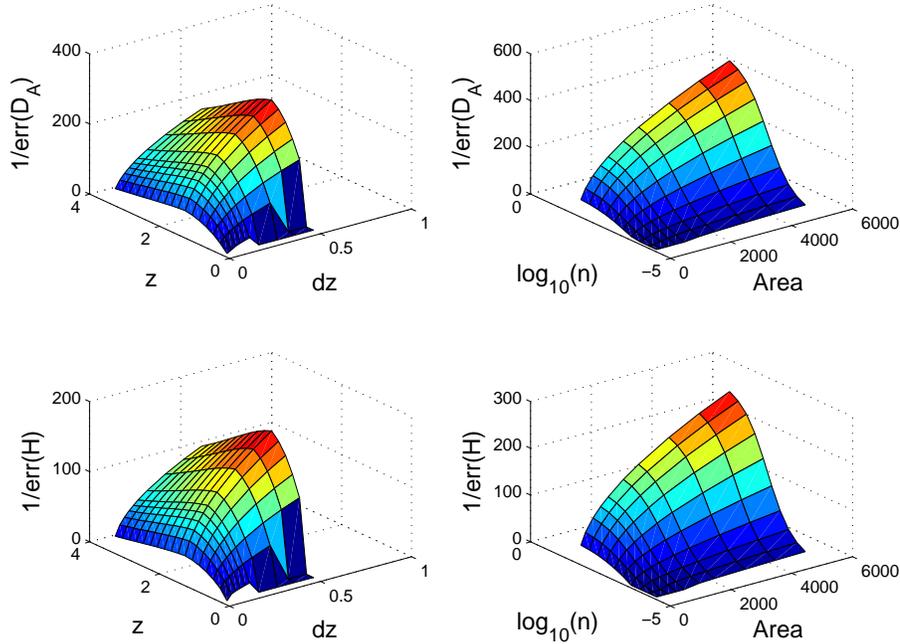,width=12cm,angle=0}
\caption{Dependence of the fitting formulae accuracies for
  spectroscopic surveys on pairs of survey parameters.  For the
  left-hand plots we vary $z$ and $\delta z$, fixing $A = 2$ and $n =
  5.1$.  For the right-hand plots we vary $A$ and $n$, fixing $z = 1$
  and $\delta z = 0.5$.  The accuracy is plotted as $1/y$.}
\label{fig3dplot}
\end{figure*}

\begin{figure}
\center
\epsfig{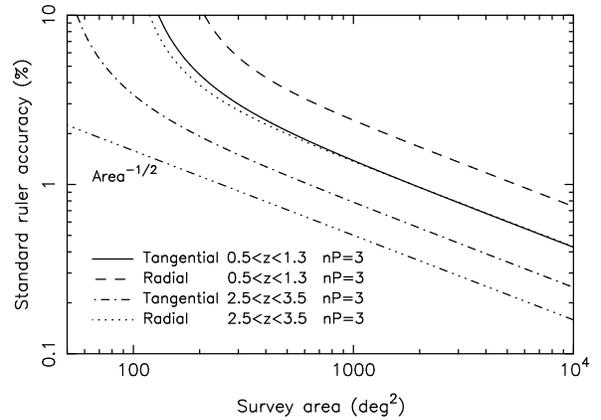}
\caption{Tangential and radial baryon oscillation accuracies for
  spectroscopic redshift surveys as a function of survey area.  We
  illustrate cases corresponding to the redshift windows that are
  naturally probed by optical spectrographs, $0.5 < z < 1.3$ and $2.5
  < z < 3.5$.  We have assumed a sufficient number density of galaxies
  that shot noise is unimportant.  A reference line Accuracy $\propto$
  Area$^{-1/2}$ is plotted.  The dependence of accuracy on area
  becomes steeper for small areas because the baryon oscillations are
  no longer being adequately resolved by the data.}
\label{figexample1}
\end{figure}

\begin{figure}
\center
\epsfig{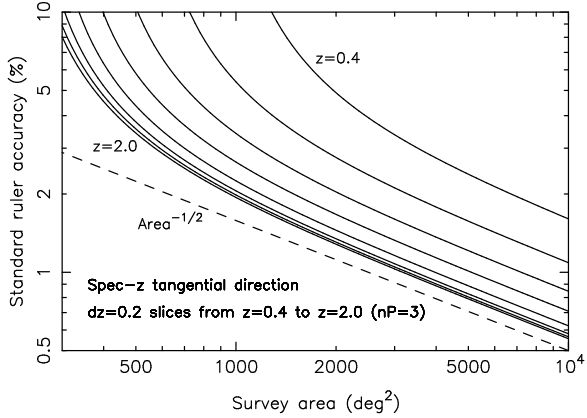}
\caption{The tangential baryon oscillation accuracy for spectroscopic
  redshift surveys as a function of survey area for a series of
  redshift slices of width $0.2$.  We have assumed a sufficient number
  density of galaxies that shot noise is unimportant.  A reference
  line Accuracy $\propto$ Area$^{-1/2}$ is plotted.  The dependence of
  accuracy on area becomes steeper for small areas because the baryon
  oscillations are no longer being adequately resolved by the data.}
\label{figexample2}
\end{figure}

\begin{figure}
\center
\epsfig{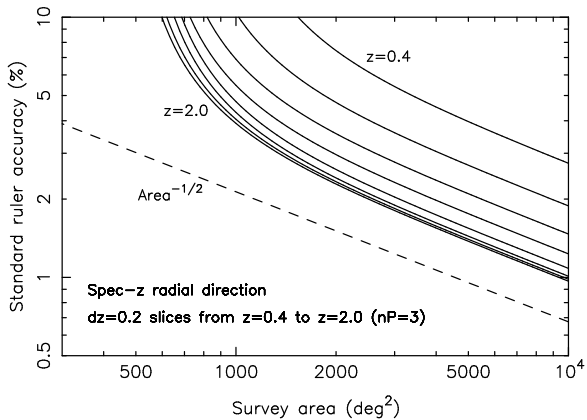}
\caption{The same as Figure \ref{figexample2}, plotting the radial
  baryon oscillation accuracy.}
\label{figexample3}
\end{figure}

\begin{figure}
\center
\epsfig{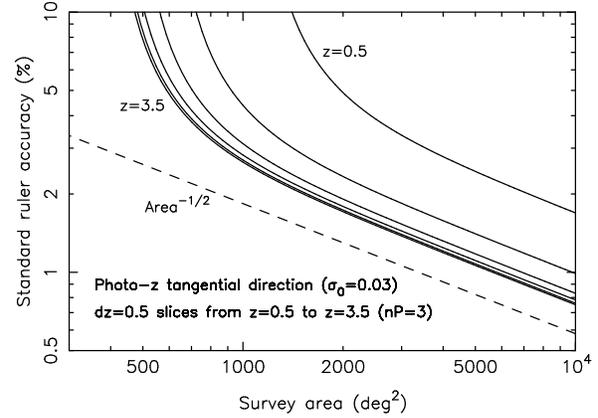}
\caption{The tangential baryon oscillation accuracy for photometric
  redshift surveys as a function of survey area for a series of
  redshift slices of width $0.5$.  We have assumed a photometric
  redshift error $\sigma_0 = 0.03$ and a sufficient number density of
  galaxies that shot noise is unimportant.  A reference line Accuracy
  $\propto$ Area$^{-1/2}$ is plotted.  The dependence of accuracy on
  area becomes steeper for small areas because the baryon oscillations
  are no longer being adequately resolved by the data.}
\label{figexample4}
\end{figure}

Figure \ref{fig3dplot} plots the fitting formulae accuracies for
spectroscopic surveys against pairs of survey parameters: $(A,n)$ and
$(z,\delta z)$.  Figures \ref{figexample1} to \ref{figexample4} plot
the accuracies against survey area for some more specific
configurations of interest.  For these last four figures we assume a
number density of objects such that shot noise is unimportant
(`$nP=3$', see Glazebrook \& Blake 2005 Figure 1 for the required
number density as a function of redshift).  Figure \ref{figexample1}
considers spectroscopic redshift surveys covering the redshift ranges
$0.5 < z < 1.3$ and $2.5 < z < 3.5$, which are naturally probed using
optical spectrographs.  Figures \ref{figexample2} and
\ref{figexample3} display tangential and radial accuracies for a more
general range of spectroscopic survey configurations in redshift
slices of thickness $0.2$ from $z = 0.4$ to $z = 2.0$.  As redshift
increases, the gain in survey volume with $z$ saturates and thus the
curves converge.  Figure \ref{figexample4} plots baryon oscillation
accuracies from photometric redshift surveys (with redshift error
parameter $\sigma_0 = 0.03$) in redshift slices of thickness $0.5$
from $z = 0.5$ to $z = 3.5$.

Comparing the predictions of the fitting formulae with results from
the full Monte Carlo method of BG03 (e.g.\ Glazebrook \& Blake 2005,
Table 1) we find that the mean difference is about $5\%$ of $y$ and
the standard deviation of the difference is roughly $10\%$ of $y$.  We
can also compare the fitting formulae prediction with the accuracy of
measurement of the acoustic scale by the SDSS LRG sample (Eisenstein
et al.\ 2005).  This survey covers sky area $3816$ deg$^2$ and
redshift range $0.16 < z < 0.47$ ($V = 0.72 \, h^{-3}$ Gpc$^3$).  The
galaxy number density varies with redshift, but we take an effective
value $n_{\rm eff} = 10^{-4} \, h^3$ Mpc$^{-3}$ and a galaxy bias
corresponding to $\sigma_8 = 1.8$ (Eisenstein et al.\ 2005).  The
fitting formulae predict measurement accuracies of $6.4\%$ ($8.5\%$)
in the tangential (radial) direction, using just the oscillatory
information.  Eisenstein et al.\ determined a $4\%$ measurement of the
acoustic scale when the clustering pattern was averaged over angles,
using the full information contained in the shape.  Our combined
tangential and radial measurements suggest an overall accuracy of
about $5\%$ from just the oscillatory component, which appears broadly
consistent.

\section{Changing the cosmological parameters}
\label{seccosmo}

These fitting formula coefficients have been derived from a grid of
simulated surveys assuming a fiducial $\Lambda$CDM cosmology.
However, the scaling arguments presented in Section \ref{secexpress}
apply more generally.  As a result, it is a good approximation to use
the fitting formula of equation \ref{eqfitform3} for a range of
cosmological parameters, if we compute the volume $V$, linear growth
factor $D(z)$ and (where applicable) the radial position error
$\sigma_r$ using the new set of parameters.  The coefficients $x_0$,
$V_0$, $\sigma_{r,0}$ and $D(z_0)$ should remain unaltered at their
$\Lambda$CDM calibrations.  However, two further changes are required:
\begin{itemize}
\item The amplitude and shape of the input power spectrum $P$ depend
  on the cosmological parameters.  Our technique is largely
  insensitive to these dependences because we divide out the overall
  power spectrum shape before fitting the baryon oscillations.
  However, the balance between cosmic variance and shot noise will be
  affected (i.e.\ the value of $nP$ in equation \ref{eqfitform1}).
  For a new set of parameters, the coefficient $n_{\rm eff}$ should be
  scaled inversely with the characteristic power spectrum amplitude
  for the scales of interest, relative to its value in the fiducial
  case.
\item We should re-estimate the cut-off redshift $z_{\rm m}$ at which
  all of the high-amplitude acoustic peaks become visible: the
  location of the non-linear transition scale at a given redshift
  depends on the growth of density perturbations, which is determined
  by the cosmological parameters.
\end{itemize}

\section{Sparse-sampling strategies}

Thus far, our formulae refer to surveys covering a fully contiguous
sky area.  However, the optimal strategy for measuring acoustic
oscillations given a fixed observing time may not be to survey a
contiguous area, but rather to sparsely-sample a larger area:
gathering a larger density-of-states in Fourier space at the expense
of an increased convolution of the input power spectrum (i.e.\ more
smoothing of the acoustic oscillations) and increased correlations
between adjacent Fourier bins (i.e.\ less statistical significance for
an observed peak or trough in power).  In practice, sparse-sampling
could be achieved by a non-contiguous pattern of telescope pointing
centres or, for a wide-field multi-object spectrograph, by
distributing the fibres non-uniformly across the field-of-view.

In the first approximation, the effectiveness of a sparse-sampling
strategy depends on the angular size $\theta$ of the observed survey
patches (e.g.\ the field-of-view of the optical spectrograph) compared
to the angular scale of the baryonic features in the power spectrum
$\sim s/r(z) = 2.6$ deg at $z = 1$.  If $\theta \sim 1$ deg then $W_k$
will contain structure on scales similar to the acoustic preferred
scale, and an unacceptable degree of convolution will result.  If
$\theta \ll 1$ deg, then a sparse-sampling strategy will usually be
preferred.

\begin{figure}
\center
\epsfig{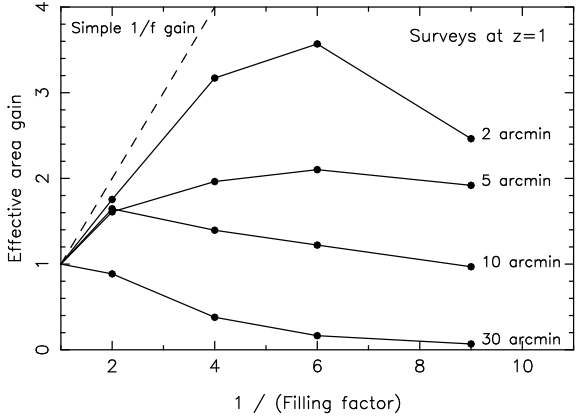}
\caption{Effective area gain for a series of sparsely-sampled survey
  strategies at $z=1$, varying the filling factor $f$ of equation
  \ref{eqfill} and the survey patch size $\theta$.  The area gains are
  evaluated using equation \ref{eqgain} and are compared with the
  simple increase in survey performance neglecting the effects of
  convolution and mode correlations (the dashed line).  The cases
  analyzed are indicated by the solid circles.}
\label{figsparse}
\end{figure}

We investigated this trade-off by simulating a series of
sparse-sampling strategies with $\theta = 2,5,10,30$ arcmin,
considering spectroscopic redshift surveys only, and measuring the
power spectrum in angle-averaged bins of constant wavenumber $k =
\sqrt{k_x^2 + k_y^2 + k_z^2}$.  For each value of $\theta$ we
considered a series of survey `filling factors' $1/f$ such that
\begin{equation}
f = \frac{{\rm Sparsely \,\, sampled \,\, area}}{{\rm Observed \,\,
    area}}
\label{eqfill}
\end{equation}
For the purposes of this simple investigation we assumed that the
survey window function was a regular grid of square patches of size
$\theta \times \theta$ (we note that other sampling strategies may be
preferred, such as a random distribution of pointings or a logarithmic
spiral).  For each $(\theta,f)$ we determined an `effective area gain'
for the sparsely-sampled survey, by which we should multiply our
observed (sparsely-sampled) area to produce the approximate input to
the baryon oscillation fitting formulae.

The Fourier transform grid required to analyze a volume large enough
to ensure a high-accuracy measurement of the baryon oscillations,
whilst maintaining a resolution several times better than the sparse
sampling scale of a few arcmin, is prohibitively large.  Therefore we
adopted a different approach, estimating the effective area gain by
quantifying three competing effects:
\begin{itemize}
\item The average decrease in the amplitude $A$ of the acoustic
  oscillations due to convolution with the window function.
\item The average decrease in the power spectrum error $\sigma$ in
  each Fourier bin (i.e.\ the diagonal elements of the covariance
  matrix) due to the increased number of Fourier modes analyzed.
\item The increased correlation of each Fourier bin with its
  neighbours, defined by quantifying an `effective number of
  independent modes' $m_i$ for each bin $i$ using the covariance
  matrix $C_{ij}$ of equation \ref{eqpkcov}.  For a uniform survey
  window function in a cuboid, $m_i = 1/(\delta P_i)^2 =
  (P+\frac{1}{n})^2/C_{ii}$, where $P$ is the power spectrum amplitude
  in bin $i$.  For a general window function we defined:
\begin{equation}
m_i = \frac{(P+\frac{1}{n})^2}{\sum_j C_{ij}}
\end{equation}
such that the off-diagonal covariance matrix elements decrease the
independence of the bins.  We take the sum over $j$ up to the
non-linear transition scale, and then define the average across the
bins, $m = \overline{m_i}$.
\end{itemize}

We initially measured these quantities for a fiducial contiguous
($f=1$) survey of $100$ deg$^2$ spanning redshift range $0.75 < z <
1.25$.  We then repeated our analysis for each pair of values of
$(\theta,f)$ defining in each case
\begin{equation}
{\rm Effective \,\, area \,\, gain} = \left(
\frac{A/\sigma}{A_0/\sigma_0} \right)^2 \frac{m}{m_0}
\label{eqgain}
\end{equation}
where the subscript `0' indicates values for the fiducial survey.  The
relative powers of the quantities are chosen in accordance with their
scaling with the number of Fourier modes $m$: Area $\propto m$ and
$\sigma \propto 1/\sqrt{m}$.  For the cases with small values of
$\theta$, the convolution involves small-scale power from the
non-linear clustering regime, thus we modified our input linear power
spectrum using the non-linear prescription of Peacock \& Dodds (1994).

The results are displayed in Figure \ref{figsparse}.  As expected,
large survey patches $\theta \ga 30$ arcmin do not favour
sparse-sampling strategies because of the consequent serious smoothing
of the acoustic oscillations.  If $\theta \la 10$ arcmin then
sparse-sampling strategies are preferred, although we note that the
resulting performance plotted in Figure \ref{figsparse}, which
includes the window function effects, is not as good as that which
would be inferred by using the entire `sparsely-sampled area' as the
input area in the fitting formula, thus neglecting the window function
effects (as indicated by the `simple area gain' line plotted in Figure
\ref{figsparse}).  We emphasize that our calculations here are only a
first approximation and this is a subject requiring further study.

\section{Summary}

We have developed a fitting formula for the accuracy with which the
characteristic baryon oscillation scale may be extracted from future
spectroscopic and photometric redshift surveys in the tangential and
radial directions, using heuristic scaling arguments calibrated using
an accelerated version of the `model-independent' method of Blake \&
Glazebrook (2003).  The formula is given in equations \ref{eqfitform3}
to \ref{eqfitform6} with the values of the parameters listed in Table
\ref{tabparam}, and reproduces the simulation results with a
fractional scatter of $7\%$ ($10\%$) in the tangential (radial)
direction, over a wide grid of survey configurations.  Simple
modifications allow the fitting formula to be applied for a range of
cosmological parameters.  We have also investigated how a simple
sparse-sampling strategy may be used to enhance the effective survey
area if the sampling scale $\theta$ is much smaller than the
characteristic angular acoustic scale ($\theta \ll 1$ deg).  This may
be implemented for a wide-field multi-object spectrograph by
clustering the fibres in the field-of-view.

\section*{Acknowledgments}

CB acknowledges current funding from the Izaak Walton Killam Memorial
Fund for Advanced Studies and the Canadian Institute for Theoretical
Astrophysics.  DP was supported by PPARC.  MK acknowledges funding
from the Swiss National Science Foundation.  RCN thanks the EU for
support via a Marie Curie Chair.  We thank Gemini for funding part of
this work via the WFMOS feasibility study.  We acknowledge the use of
multiprocessor machines at the ICG, University of Portsmouth.

\end{document}